\documentclass[myepj-spec,pdftex]{mySvjour}
\usepackage{graphicx}
\usepackage{hyperref}
\usepackage{color}
\usepackage{xspace}
\usepackage[square,sort&compress,numbers]{natbib}   
\usepackage{verbatim}
\usepackage{amsmath,amssymb,amsfonts} 
\usepackage{tabularx}
\usepackage{multicol}
\usepackage{footnote}
\usepackage{listings}
\usepackage[switch, modulo]{lineno}

\usepackage{lmodern,bm}                
\usepackage[T1]{sansmath} 
\SetMathAlphabet{\mathsfbf}{sans}{\sansmathencoding}{\sfdefault}{bx}{sl}
\usepackage{etoolbox}
\AtBeginEnvironment{sansmath}{}{}{}

\usepackage{lipsum}

\newcommand{\mueV}{\mu\text{eV}}
\newcommand{\GHz}{\text{GHz}}

\definecolor{darkblue1}{rgb}{0,0,.2}
\definecolor{darkblue}{rgb}{0,0,.2}
\definecolor{darkred}{rgb}{0.5,0,0}
\pagecolor{white} 
\color{black}     
\hypersetup{breaklinks=true, 
	colorlinks=true, 
	linkcolor=darkblue1, 
	menucolor=darkblue1, 
	urlcolor=darkblue1,
	citecolor=darkblue1,
	pdftitle={},
	pdfauthor={},
	pdfsubject={},
	pdfkeywords={},
	pdfproducer={}
}
\usepackage{siunitx}

\bibstyle{plain}
\newcommand{\bi}{\begin{itemize}}
\newcommand{\ei}{\end{itemize}}

\newcommand{\ben}{\begin{enumerate}}
\newcommand{\een}{\end{enumerate}} 

\newcommand{\bt}[1]{\begin{table}[tb]\begin{tabular}{#1} \hline\hline  \\[-1.0em]}
\newcommand{\et}[2]{\hline\hline \end{tabular} \caption{#1} \label{#2} \end{table}}

\newcommand{\be}{\begin{equation}}
\newcommand{\ee}{\end{equation}}

\newcommand{\bea}{\begin{eqnarray}}
\newcommand{\eea}{\end{eqnarray}}

\newcommand{\bc}{}

\newcommand{\mev}{\ensuremath{\mathrm{\,Me\kern -0.1em V}}\xspace}
\newcommand{\gev}{\ensuremath{\mathrm{\,Ge\kern -0.1em V}}\xspace}

\begin{document}
	
	\twocolumn[{%
		\begin{@twocolumnfalse}
			
			\begin{flushright}
				\normalsize
			\end{flushright}
			
			\vspace{-2cm}
			\title{\Large\boldmath The Global Network of Cavities to Search for Gravitational Waves (GravNet): A novel scheme to hunt  gravitational waves signatures from the early universe}
			%

\author{Kristof Schmieden \inst{1} \and Matthias Schott \inst{1,2}}
\institute{\inst{1} PRISMA+ Cluster of Excellence, Institute of Physics, Johannes Gutenberg University, Mainz, Germany, \\ \inst{2} Department of Physics, Stony Brook University, USA}


			\abstract{The idea of searching for gravitational waves using cavities in strong magnetic fields has recently received significant attention. Specifically, discussions focus on cavities with relatively small volumes, which are currently employed in the search for axions. 
			In this context, we propose a novel experimental scheme that enables the detection of gravitational waves in the GHz regime, which could originate, for example, from primordial black hole mergers. 
			The scheme is based on synchronous measurements of cavity signals from multiple devices operating in magnetic fields at distant locations. 
			While gravitational wave signatures might be detectable in individual cavities, distinguishing them from noise is highly challenging. 
			By analyzing the correlation among signals from several, possibly geographically separated cavities, it is not only possible to significantly enhance the signal-to-noise ratio but also to investigate the source of those gravitational wave signatures. 
			In the context of this proposal, a first demonstration experiment with one superconducting cavity is currently conducted, which is the basis of the proposed data-analysis approaches. On this basis the prospects of GravNet (Global Network of Cavities to Search for Gravitational Waves) are outlined in the paper.
			}	
	\maketitle
	\end{@twocolumnfalse}
}]

\tableofcontents

\section{Introduction}	

The detection of gravitational waves (GWs) by the LIGO and Virgo interferometers \cite{LIGOScientific:2016aoc} marked the beginning of a new era in astronomy. 
Gravitational waves, with frequencies spanning from super massive binary black hole systems in the nHz regime to kHz for compact binary objects and up to GHz for GWs from the cosmic gravitational wave background \cite{Aggarwal:2020olq}, are an essential part of our understanding of the universe.

Interferometers, like LIGO and Virgo, have proven to be highly successful in detecting GWs, and future generations, such as the Einstein Telescope \cite{Punturo:2010zz}, are in the design phase. 
An alternative concept for GW detection exploits their coupling to the electromagnetic field, using radio frequency cavities, either pumped or placed in a static magnetic field. Recently, the latter approach has been discussed in more detail \cite{Berlin:2021txa, Domcke:2022rgu, Ringwald:2020ist}, especially in the context of searches for axion-like particles \cite{ADMX:2021nhd, Schmieden:2021msb, Berlin:2022hfx}.

The basic principle behind the cavity-based experiments is simple: a gravitational wave distorts the cavity's shape, altering the magnetic flux through the cavity and generating an electric signal that can be detected. Additional, the GW couples directly to the EM field via the inverse Gertsenshtein effect.
Hence, a gravitational wave that is passing through a cavity within a static magnetic field, creates an effective current in Maxwell's equations, leading to an electromagnetic field that oscillates at the same frequency as the gravitational wave. The induced electromagnetic field can be resonantly enhanced using microwave cavities and the generated radio frequency power detected.

The sensitivity of such experiments depends on the GW frequency, incoming direction, the cavity's resonance frequencies, and the external magnetic field strength. The sensitivity to gravitational waves using a cavity-based experiment has been derived in \cite{Berlin:2021txa} and can be summarised by the signal power
\begin{equation}\label{eq:signalPower}
P_{sig} = \frac{1}{2} Q \omega_g^3 V^{5/3} (\eta _n h_0 B_0)^2 \frac{1}{\mu_0 c^2},
\end{equation}
with $\omega_g$ denoting the GW frequency and $h_0$ the magnitude of the GW strain. The cavity is described by its volume $V$, its quality factor $Q$ as well as the external magnetic field $B_0$. The dimensionless coupling constant $\eta _n$ is given by the overlap of the induced current with the excited cavity mode:
\begin{equation}
\eta_n = \frac{|\int_{V} d^3x E^*_n \cdot \hat j_{+,\times}|}{V^{1/2} (\int _V d^3x |E_n|^2)^{1/2}},
\end{equation}
where $E_n$ denotes a resonant mode of the cavity and $\hat j_{+,\times}$ describe spatially-dependent dimensionless functions for the spatial profile and polarization of the GW signal. We refer to \cite{Berlin:2021txa} for further details. 
It is important to note that the sensitivity increases quadratically with the magnetic field strength $B_0$ and to the power of 5/3 with the cavity volume V. 

Unlike axion searches, where each axion mass corresponds to a single resonance frequency, GW signals may be broadband and not localized at a specific frequency. 
Even more importantly, GW are expected to be coherent over large distances, in contrast to axion halo signatures. 
This opens up a completely new approach to search for GW: instead of constructing one dedicated cavity based experiment, a global network of several cavity experiments over the whole world could coherently combine their measurements and boost significantly the sensitivity. 
In particular, it is sufficient to use off-shelf high field magnet systems, which can be also operated at smaller laboratories. 

We present in the following the basic idea of a Global Network of Cavities to Search for Gravitational Waves (GravNet). After a brief discussion on the potential sources of the high frequency gravitational waves (Section \ref{sec:GWsources}), we briefly discuss the baseline cavity experiment SUPAX (Section \ref{sec:concept}), which will be used to estimate the overall baseline sensitivity of GravNet. Extensions of GravNet are summarized in Section \ref{sec:gravNet-1}. 

\section{Sources of high frequency gravitational waves}
\label{sec:GWsources}

Various sources have the potential to produce high-frequency gravitational waves (GWs), and they have been studied extensively \cite{Berlin:2021txa}. Notably, mergers involving sub-solar mass objects, including primordial black holes (PBHs) and other exotic compact objects, are typical examples. Additionally, GWs can be generated by boson clouds via black hole superradiance. Also exotic compact objects, such as boson and fermion stars, gravitino stars, gravistars, and dark matter blobs, can have significantly lower masses than a solar mass, enabling them to emit GWs at high frequencies. 

During the inspiral phase of a binary merger, the frequency of the emitted GWs increases as the compact objects approach each other. There exists an upper bound on the GW frequency emitted by a binary in a quasi-circular orbit, which corresponds to the frequency at the innermost stable circular orbit (ISCO). For a binary with equal-mass compact objects ($M_b$) during the early stages of the merger, the GW frequency evolves as:
\[ \omega_g \approx 14 \, \text{GHz} \times \left(\frac{10^{-6}M_\odot}{M_b}\right)\left(\frac{r_{\text{ISCO}}}{r_b}\right)^{3/2}, \]
where $M_\odot$ denotes the solar mass. Therefore, only very light binaries, such as sub-Earth mass PBHs, can generate GW signals in the GHz regime well before reaching the ISCO.

However, GW signals from such light binaries become highly transient near the ISCO, as the emitted GW frequency evolves over time according to:
\[ \frac{d\omega_g}{dt} \approx \left(\frac{M_b}{r_b}\right)^{11/6}. \]
This time dependence limits the sensitivity of resonant experiments. The number of orbital cycles ($N_{\text{cyc}}$) a binary spends emitting GWs within the resonator bandwidth $\omega_g/Q$ for frequencies in the GHz regime can be estimated as:
\[ N_{\text{cyc}} \approx 10^{-3} \times \left(\frac{10^6 M_\odot}{M_b}\right)^{5/3} \left(\frac{10^5}{Q}\right). \]
Requiring at least $N_{\text{cyc}}>10^5$ for the detection of a GW signal implies $M_b < 10^{-11} M_\odot$. Remarkably, this is the regime where PBHs could constitute a significant fraction of the cosmological dark matter abundance \cite{Berlin:2021txa}.
The dependence of the time the PBH merger signal spends within a cavity bandwidth is shown in Figure \ref{fig:h0_intTime}. 

\begin{figure*}
\centering
\includegraphics[width=8.5cm]{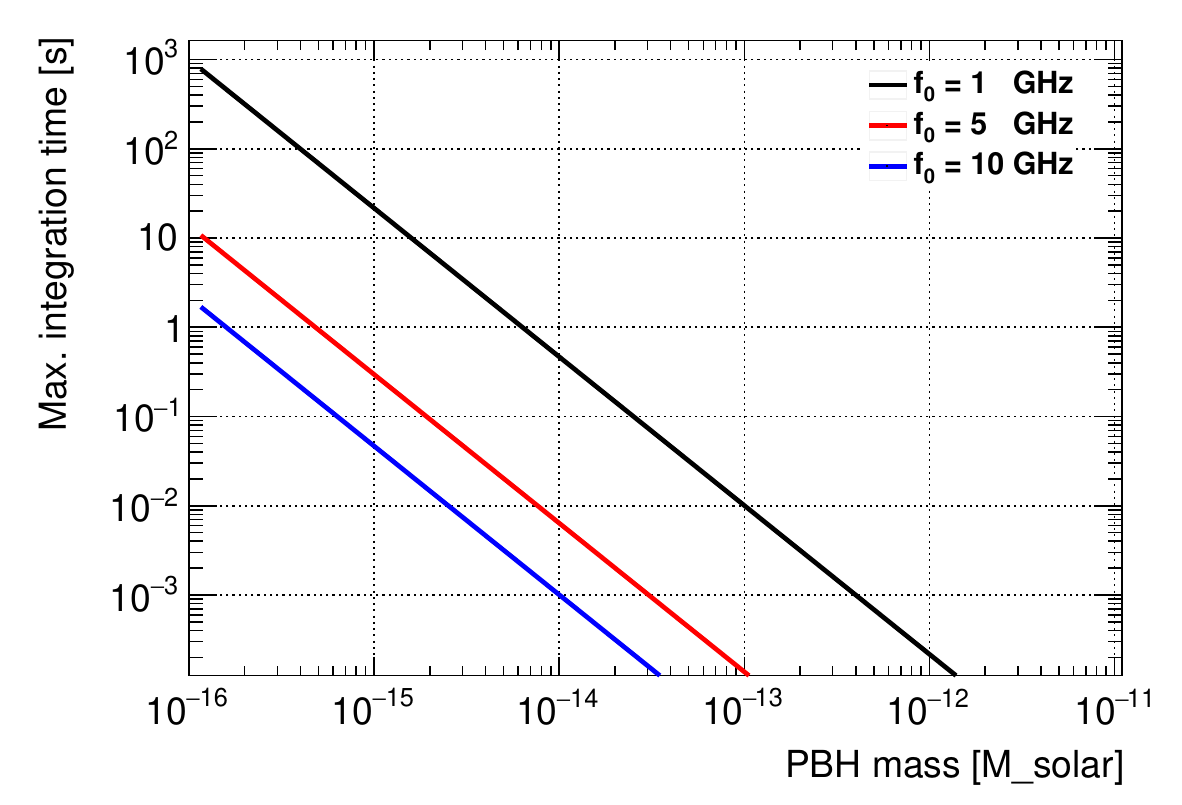}
\includegraphics[width=8.5cm]{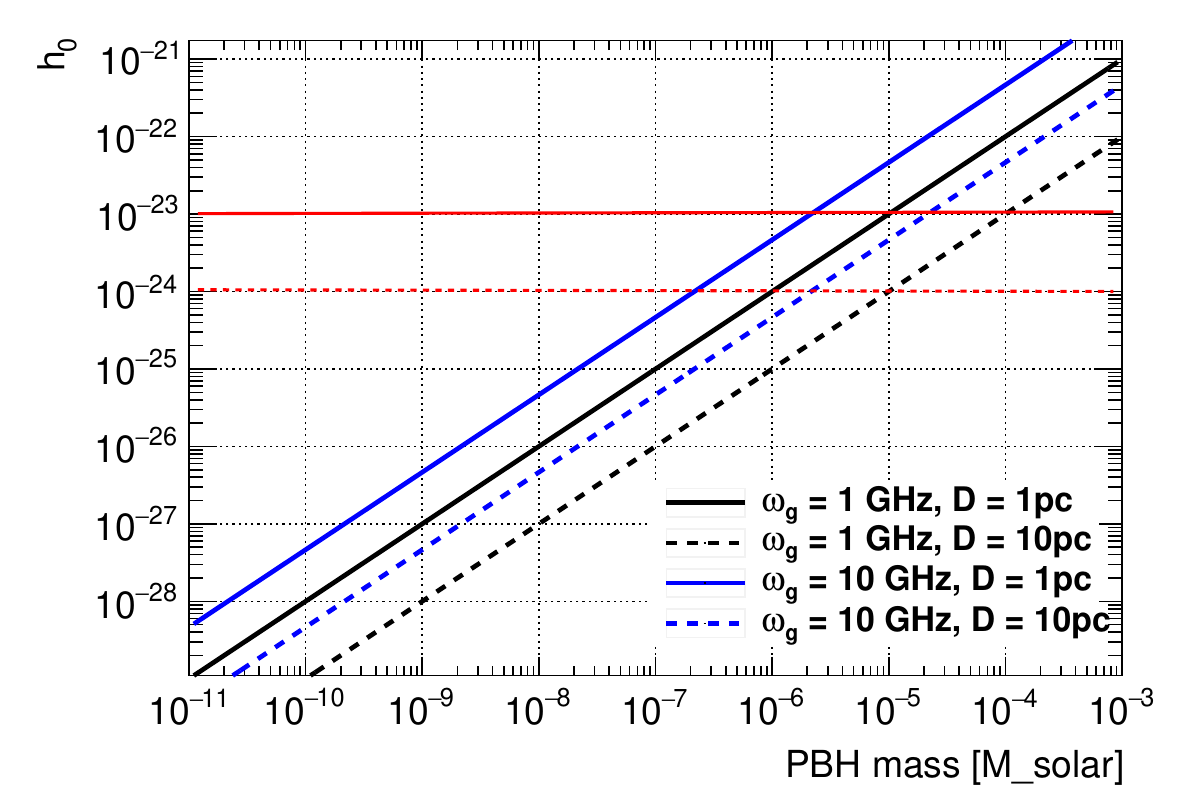}
\caption{Left: Maximum feasible integration time given by the duration the frequency change is less than $f_0/Q$, assuming $Q = 10^6$, for a PBH merger. Right: typical strains of PBH mergers in dependence of their mass for distances of 1pc and 10pc from the earth.}\label{fig:PBHMergerStrain}
\end{figure*}
However, other sources such as GWs from superradiance of boson clouds in the vicinity of PBHs and other more exotic objects are non-transient and remain at constant frequencies over long time intervals, order of months, allowing for long measurement times \cite{Arvanitaki:2009fg}. 

The typical distance of high frequency gravitational wave sources is in the order of 1 pc or more. 
Hence the GW wavefront at earth can be assumed to be coherent across distances at earth, in particular when the origin of the wave can be determined, as the phase-differences at different experimental locations at earth can be calculated. 

\section{Basic Considerations on Cavity Designs\label{sec:concept}} 	

In order to estimate the expected signal over noise ratio for a typical cavity of the GravNet proposal, we use the existing numbers of the SUPAX experiment, which is currently in operation at the University of Mainz \cite{Schneemann:2023bqc}. The SUPAX cavity and its readout will be discussed in section \ref{sec:supax}. 
We then discuss the advantages of combining several cavities in a coherent manner in section \ref{sec:combiningCavities}, before we discuss single photon detection approaches in cavities in section \ref{sec:SPC}. 

\subsection{The SUPAX Cavity Design \label{sec:supax}}

SUPAX is an experiment which aims for the search for dark photons and axion-like particles with masses around $34\,\mueV$ using a cavity in a high magnetic field \cite{Schneemann:2023bqc}. 
\begin{figure*}
\centering
\begin{minipage}[b]{5.4cm}
\centering
\includegraphics[height=6.5cm]{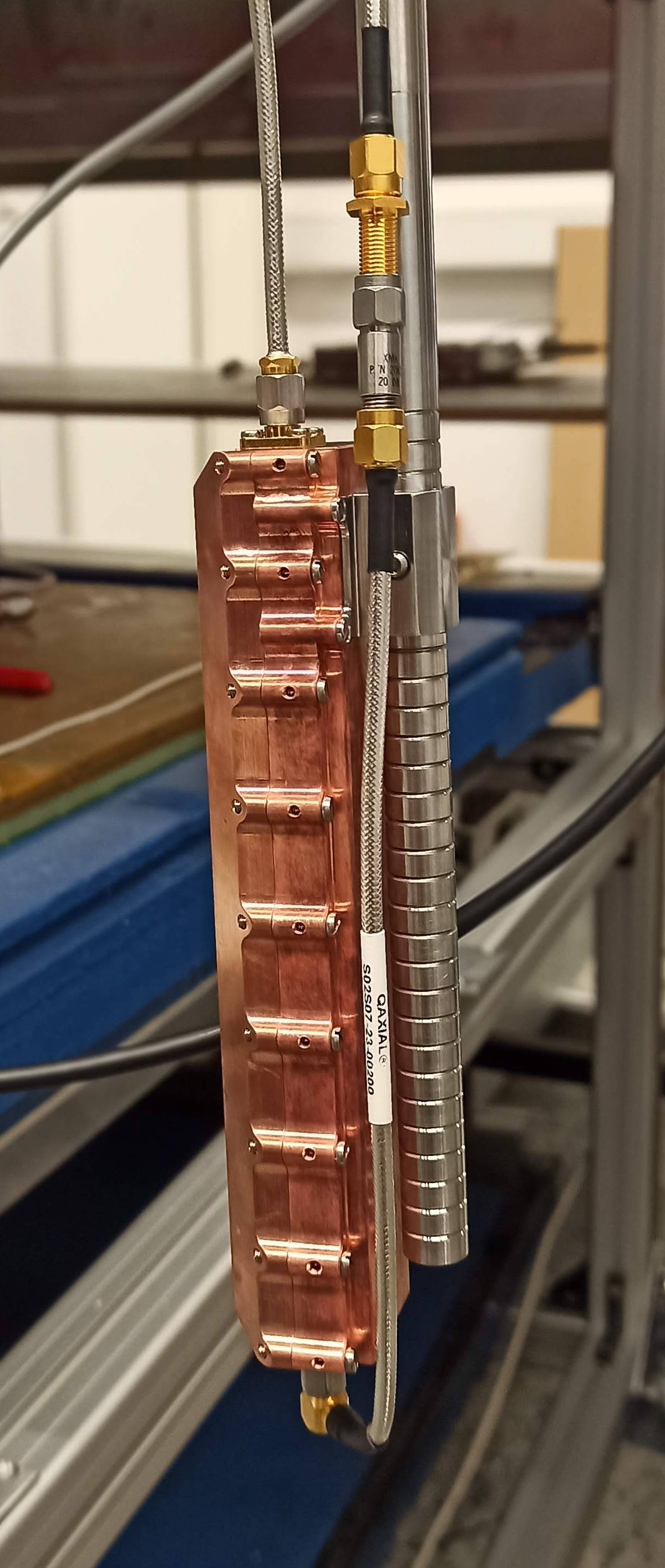}
\caption{Cavity of the SUPAX experiment with a length of 16cm.}\label{fig:supax_cavity}
\end{minipage}
\hspace{0.2cm}
\begin{minipage}[b]{10.9cm}
\centering
\includegraphics[height=8cm]{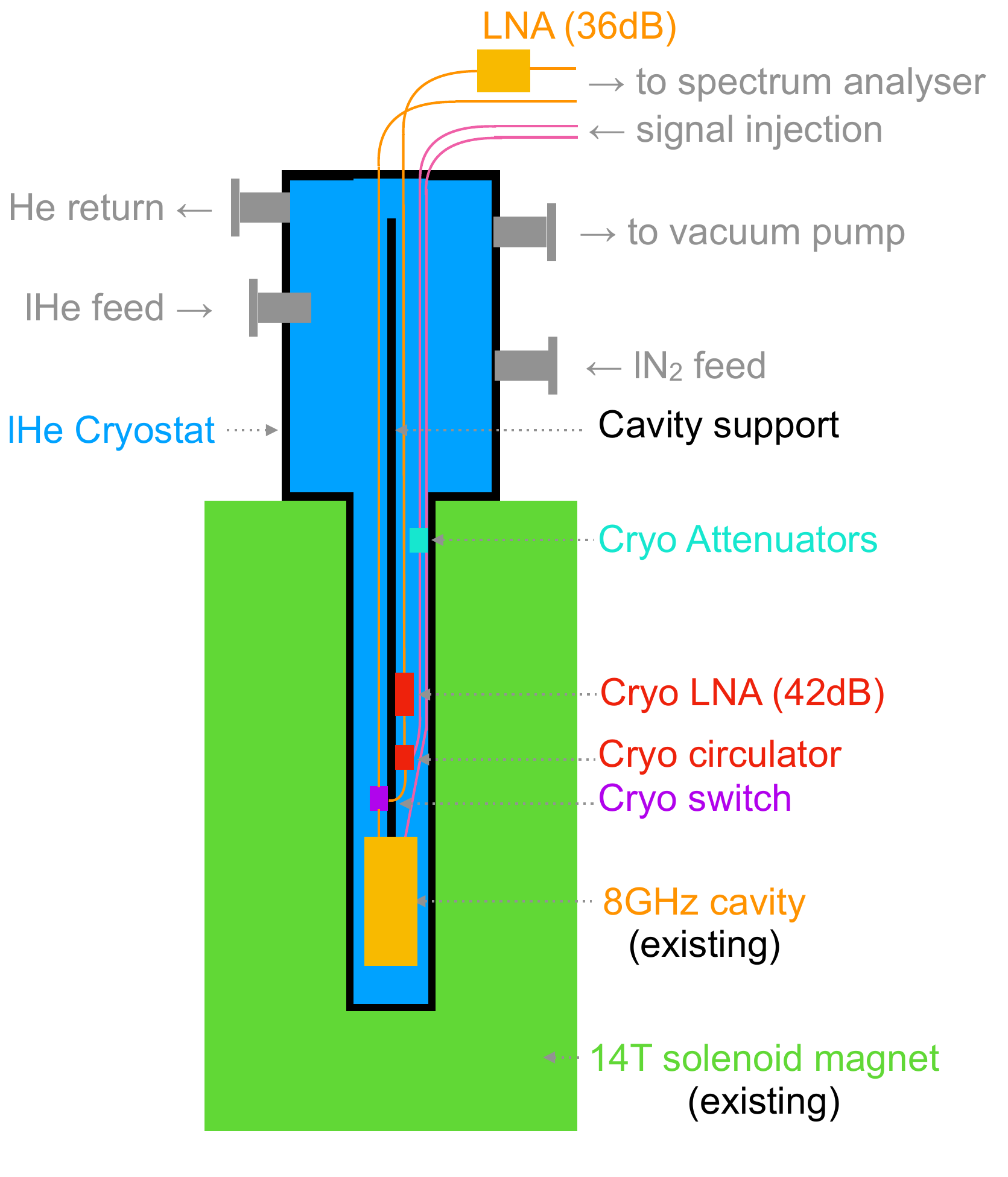}
\caption{Schematic overview of the SUPAX experimental setup.}\label{fig:SupaxSchematic}
\end{minipage}
\end{figure*}
SUPAX uses a rectangular cavity which is suspended from the top of a liquid helium (LHe) dewar and immersed in the LHe bath, as illustrated in Figure \ref{fig:SupaxSchematic}. The dimensions of the cavity are carefully chosen to maximize its effective volume while ensuring that the final assembly fits within the designated cryostat, which has a diameter of \SI{52}{mm}. Additionally, the mode next to the investigated $\text{TM}_{010}$ mode is kept at least $>\SI{100}{MHz}$ away. 
This leads to inner dimensions of the cavity of $150 \times 22.8 \times 30 \si{mm^3}$.
These dimensions yield a resonance frequency of:
\[ f = \SI{8.47}{GHz} \hspace{2mm} \Rightarrow \hspace{2mm} m_{A'} = 2 \pi \hbar \cdot f \approx \SI{35}{\mu eV}. \]
at room temperature. The cavity depicted in Figure \ref{fig:supax_cavity}, was milled from a solid copper block. 

%
The unloaded quality factor is measured to be $Q_0 = 40726 \pm 4250$ in good agreement with the simulation result of $Q_0^{\text{sim}} = 40059 \pm 128$.

The readout is based on a real-time spectrum analyser which is capable of streaming IQ time-series data with a bandwidth of 40MHz around the 8 GHz centre frequency to a readout computer. For the purpose of the axion (dark photon) search the data is immediately converted into the frequency domain and the  
resulting spectra stored for offline analysis. 
To study the offline combination of multiple cavity signals the IQ data is stored. 


The next generation cavity of the SUPAX experiment is coated with niobium-nitrite superconductor. 
This will lead to an expected improvement of the quality factor by a factor of 10 to 100. The RF behaviour of the coating in the magnetic field is currently under investigation.

The magnet to be used is a fast-ramping superconducting solenoid from CRYOGENIC LTD with a central field of up to 14T over a length of 20cm\footnote{The magnet is currently installed at the Helmholtz institute at Mainz by the group of Prof. Budker}. The room temperature bore has a diameter of $89\,\text{mm}$ which limits the diameter of the used cryostat.

\subsection{Combination of several cavities with a coherent signal}
\label{sec:combiningCavities}
Typical distances of sources of high frequency gravitational waves are 1 pc or more, hence their sources can be thought of point-like at earth, yielding coherent GW wave-fronts. 

\begin{figure*}
\centering
\includegraphics[height=3.0cm]{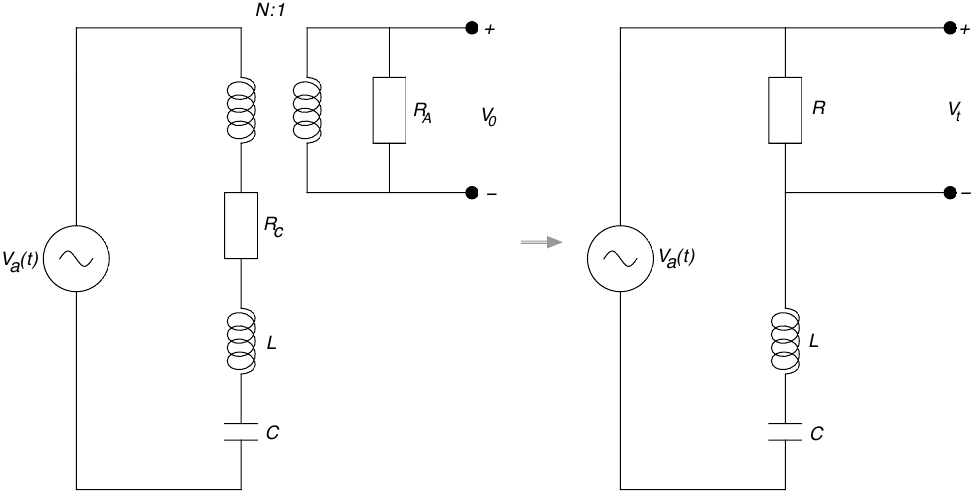}
\hspace{0.2cm}
\includegraphics[height=3.0cm]{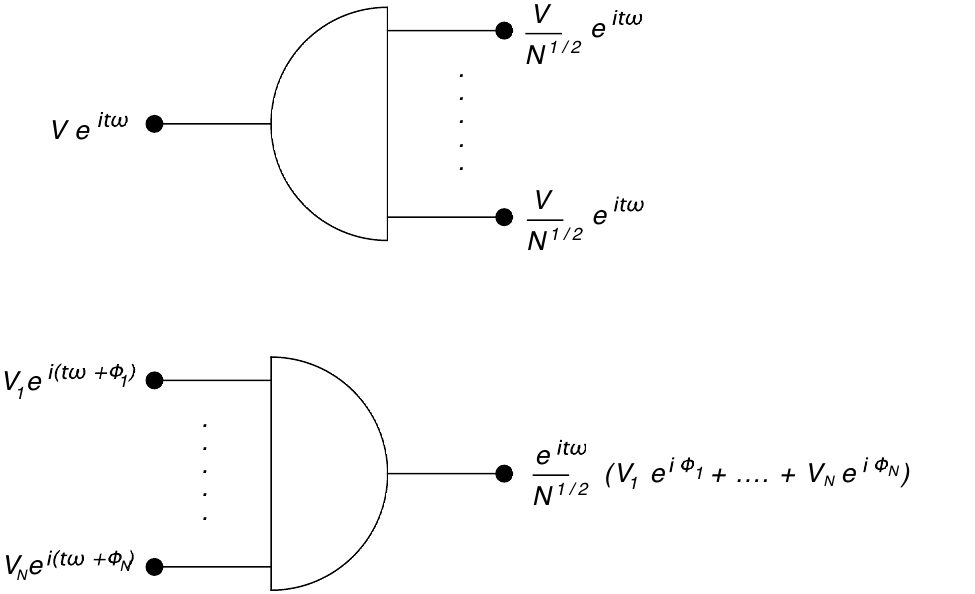}
\hspace{0.2cm}
\includegraphics[height=2.9cm]{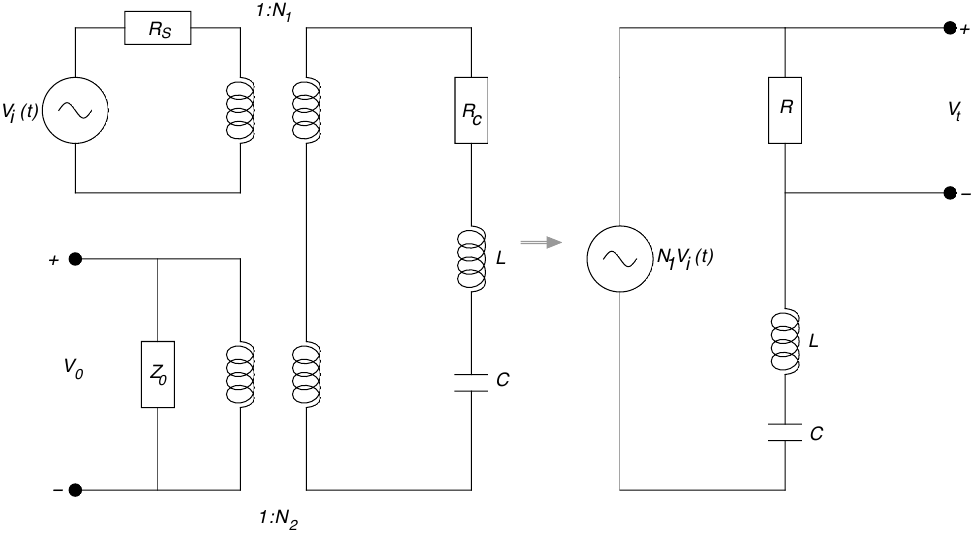}
\caption{Left: Equivalent circuit for one mode of a single-port cavity, Middle: Schematic of an N-Port Power Combiner / Divider, Right: Equivalent circuit for one mode of a two-port cavity}\label{fig:NCavities}
\end{figure*}

Before discussing the combination of locally separated cavities, we discuss first the combination of N cavity signals, recorded at the same location. 
This gives a factor of N improvement in the SNR over the individual cavity, and not as trivially assumed an improvement of $1/\sqrt{N}$. To understand this feature, we follow the discussion in \cite{ThesisKinion:1994} and consider an N-port power combiner/divider as in Figure \ref{fig:NCavities}. 
The middle picture shows how N voltages with frequency $\omega$ and phase differences $\phi$ combine. Assuming the same phase and amplitude at each input the output Voltage behaves like $V_{out} = \sqrt{N} V_{in}$, hence the power scales linearly with $N$. 

%
%
%

Having $N$ cavities distributed across earth, one needs at least three cavities which define the incoming direction of the gravitational wave. This defines the phase-difference at any further cavity, thus yielding an increase on the SNR by $N-3$ for $N$ similar cavities that are distributed across earth. 
As one single setup will not be sensitive enough to reliably detect a GW, the direction of the incoming GW, and hence the phase alignment of the various setups, will be stepped through all possible directions in the offline data analysis. 

\subsection{Single Photon Counting in Cavities \label{sec:SPC}}

Single photon counting in cavities is a cutting-edge technique that has revolutionized the field of quantum optics and precision measurements. 
It involves detecting and quantifying individual  photons that are confined within cavities. 
For RF photons the problem is much more difficult than for optical wavelengths as the energy per photon is lower. However, recently sensitivity to single RF photons was experimentally shown and several further techniques are promising to reach single RF photon sensitivity soon. 

Among the currently discussed detection methods are 
 current biased Josephson Junctions (CBJJ) \cite{DElia:2023baj}, Kerr Josephson parametric amplifiers \cite{Petrovnin:2023hho} and transmon Q-bits \cite{Balembois:2023bvs}.
While CBJJs do not yet reach single photon sensitivity, the latter two approaches do with the transmon Q-bit showing a detection efficiency of $43\%$ for $7\,\text{GHz}$ photons at a dark count rate below $90\,Hz$.

Similar to the resonant cavity readout, one would also expect a huge advantage when combining several cavities, which are distributed across the earth. 
Assuming a single photon detection efficiency of 50\%
and an improved background rate of 10Hz for one cavity one would get $3\times10^8$ background events per year, while a signal is expected only once per year. 
Since the background is uncorrelated between independent cavities, one can require a coincidence signal across N cavities. Assuming a timing resolution of 0.2 ms the background is suppressed by a factor of $(1/0.2,\text{ms}\times 10\,Hz)^{N-1}=500^{N-1}$. 
Requiring at least N=5 cavities to record a coincident signal suppresses the background therefore by $\approx 6 \times 10^{10}$, i.e. yielding a background free search. 
The problem is that 5 coincident cavities also have a resulting signal efficiency of $0.5^5\approx 0.03$ corresponding to an average time for one recorded signal event of 32 years. However, if one combines e.g. $N>5$ cavities, the likelihood to record a signal in at least 5 cavities increases according to a binomial likelihood distribution. 
Clearly, at least three cavities are required to define the direction of the incoming GW, i.e. to define the coincidence window for any other cavity. 
Assuming $N=20$, of which $N=17$ are actually available for the analysis, and a single signal efficiency of $p=0.5$, a total signal efficiency of $>90\%$ is be expected.
The optimisation of the coincidence setup is detailed in section \ref{sec:gravNet-2}.

\section{GravNet as Resonant Cavity Experiment (GravNet-1)}
\label{sec:gravNet-1}

The first implementation of GravNet will most likely be based on the usage of the resonant cavity approach, as the underlying technologies are all well tested and understood.

\subsection{Setup}

As discussed in the previous section, typical commercially available (and therefore cost-efficient) high-field magnets have a cylindrical volume of a high constant magnetic field. 
Typical dimensions range between radii of 2 to $5\,\text{cm}$ with a height of 10 to $40\,\text{cm}$. 
The coupling constant of GW signals to the cavity is in the order of $\approx 0.1$ for cylindrical cavities, but $1$ for spherical cavities. 
Assuming a constant magnetic field in a cylindrical volume of $r=4\,\text{cm}$ and $h=24\,\text{cm}$, one can either fit one cylindrical cavity with those dimensions or three spherical cavities of $r=4\,\text{cm}$. 
Comparing the effect on $P_{sig}$ between one cylindrical and one spherical cavity, this implies a factor on $P^{\text{cylinder}}_{sig}/P^{\text{spherical}}_{sig}$ of $\approx 12$ due to larger volume of the cylinder, but a factor of $\eta^2=0.01$ due to the coupling. 
The overall effect on $P_{sig}$ is roughly up to a factor $10$ in favor of a spherical cavity. Using three coupled cavities instead will increase the effective volume three fold and the SNR by a factor of 6, as can be seen from eq. \ref{eq:signalPower}. 

According to those considerations, we foresee three cavities for one experimental setup of GravNet, each with a radius of $4\,\text{cm}$, placed in a constant magnetic field of 14 T with a height of $24\,\text{cm}$.  
Finally, we assume that the readout system for GravNet will be based on Josephson Parametric Amplifiers (JPAs) or similar low noise amplifiers 
with system temperatures of $0.1\,\text{K}$.

An overview of all critical parameters of one experimental GravNet setup is given in Table \ref{tab:GarvNat1_parameters}. 

\begin{table}[ht]
    \centering
    \begin{tabular}{c|c|c}
        Setup & Supax & GravNet \\
        \hline 
        Shape   & cyl. & spher. \\
        $f_0$ [GHz] & 8.3 & 5.0  \\
        Volume [l] &  $0.128$ & $0.21$ \\
        $Q_0$ &  39600 & $10^6$ \\
        $\eta$ & 0.08 & 0.6 \\
        $T_{\text{sys}}$ [K] & 5 & 0.1  \\
        $B$ [T] & \multicolumn{2}{c}{14}  \\
        int. time & \multicolumn{2}{c}{1 s} \\
        n cavities & 1 & 3 \\
        \hline
        noise power [W] & $1.5\cdot10^{-21}\,W$ & $6.2\cdot 10^{-23}\,W$ \\
        $h_0 (P_{\text{sig}} = P_{\text{noise}})$ & $7.1\cdot10^{-21}$ & $5.2\cdot10^{-23}$ \\
    \end{tabular}
    \caption{Parameters of the experimental setup defining the signal and noise power. The measured values were obtained using the Supax Cu cavity in LHe. The expected values assume a superconducting, spherical cavity with 4 cm radius.}
    \label{tab:GarvNat1_parameters}
\end{table}


\subsection{Sensitivities}

We think it is realistic to assume that the final GravNet experiment will combine $N=10$ individual experimental locations across the globe, each hosting three cavities as detailed in the previous section. 
Assuming 1s integration time and the sensitivity for each setup as listed in Table \ref{tab:GarvNat1_parameters}, the sensitivity on $h_0$ will improve by a factor $\sqrt{N}$ to $h_0 < 1.7\cdot10^{-23}$. This requires a phase aligned combination of the time-series data of each of the setups, yielding a linear increase in the 
SNR with increasing number of setups. To this end the direction of the incoming 
GW has to be known to be able to calculate the relative phase differences between the setups depending on their geographic location. This can be achieved by assuming any direction of the GW, combining the data and searching for signal and then scanning through both angles defining the direction of the GW. 

The sensitivity can be trivially increased by increasing the integration time of the cavity signals. While integration times over hours are feasible, this requires a GW source with stable frequency over at least the intended integration time. 
Bosonic clouds exhibiting gravitational superradiance a superb candidate. 

\begin{figure}
    \centering
    \includegraphics[width=0.48\textwidth]{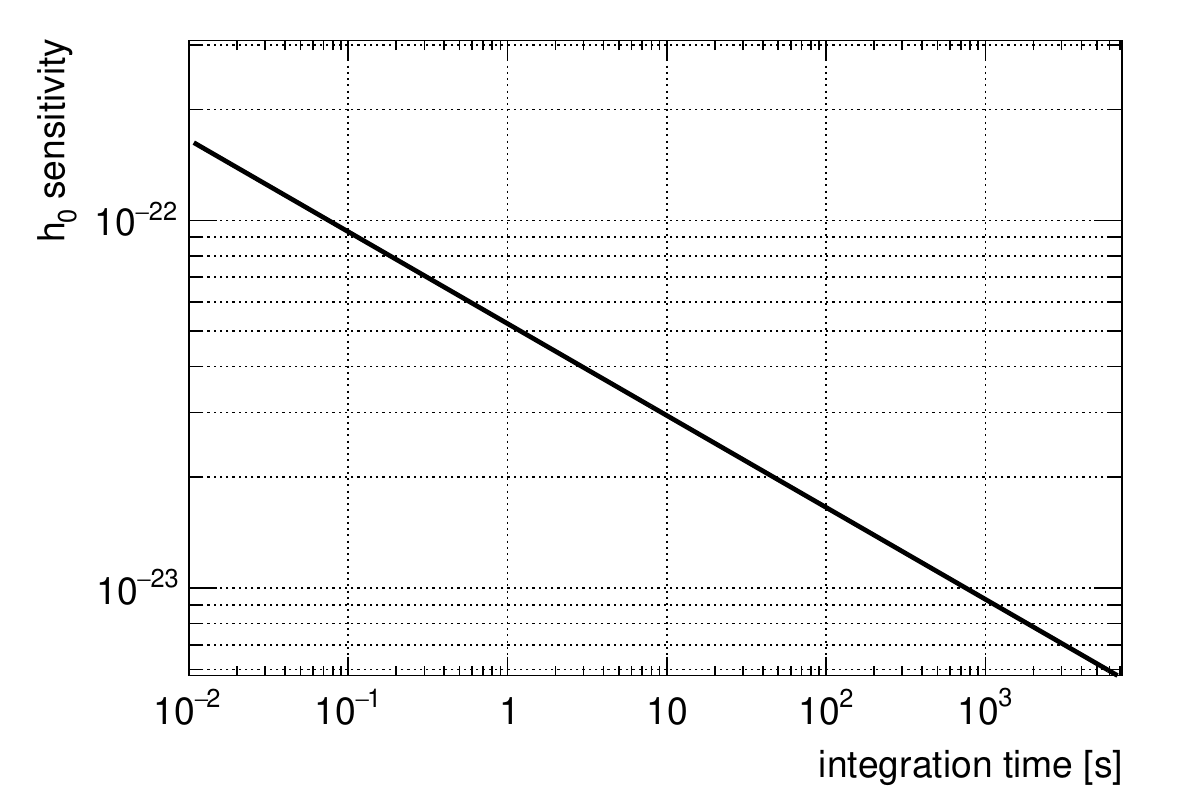}
    \caption{Shown in the sensitivity on the GW strain $h_0$ in dependence on the integration time for the resonant cavity setup with parameters assumed as shown in Table \ref{tab:GarvNat1_parameters}.}
    \label{fig:h0_intTime}
\end{figure}

Integrating for two hours instead of 1 second leads to an improved by one order of magnitude on the GW strain, reaching $h_0 < 5.6\cdot10^{-24}$ for a single setup and consequently about $1.7 \cdot10^{-24}$ for ten setups. The dependence of the sensitivity on $h_0$ on the integration time is shown in Figure \ref{fig:h0_intTime}.

\begin{figure*}
\centering
\begin{minipage}[b]{7.4cm}
\includegraphics[width=7.3cm]{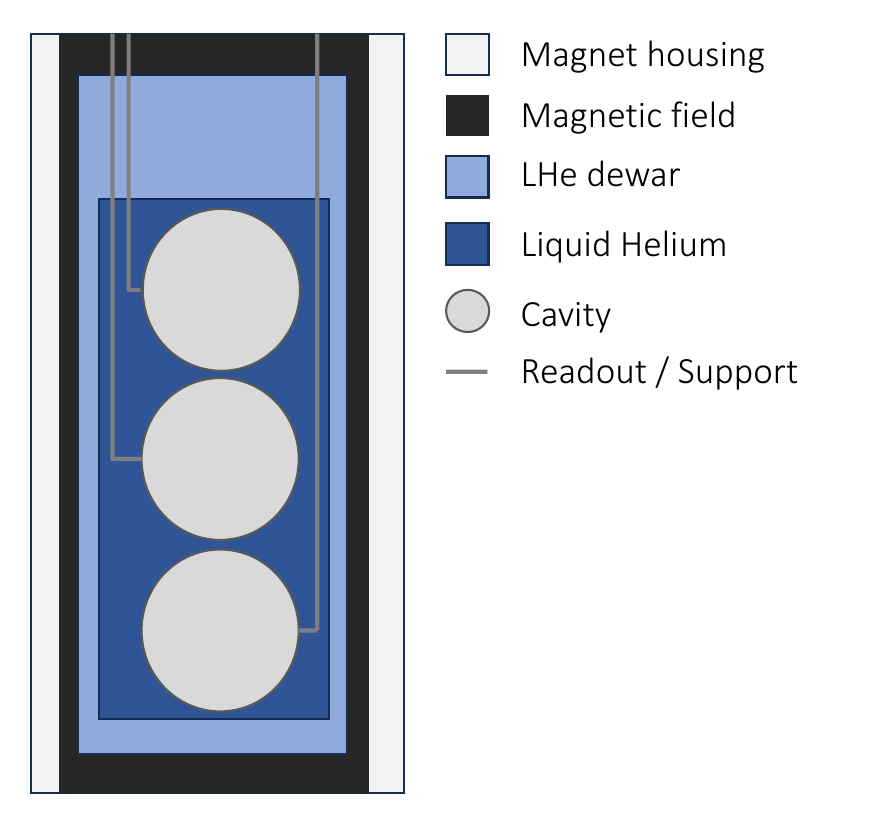}
\caption{Schematic drawing one experimental GravNet-1 setup with resonant cavities.}\label{fig:setup_scheme1}
\end{minipage}
\hspace{0.2cm}
\begin{minipage}[b]{7.4cm}
\includegraphics[width=7.3cm]{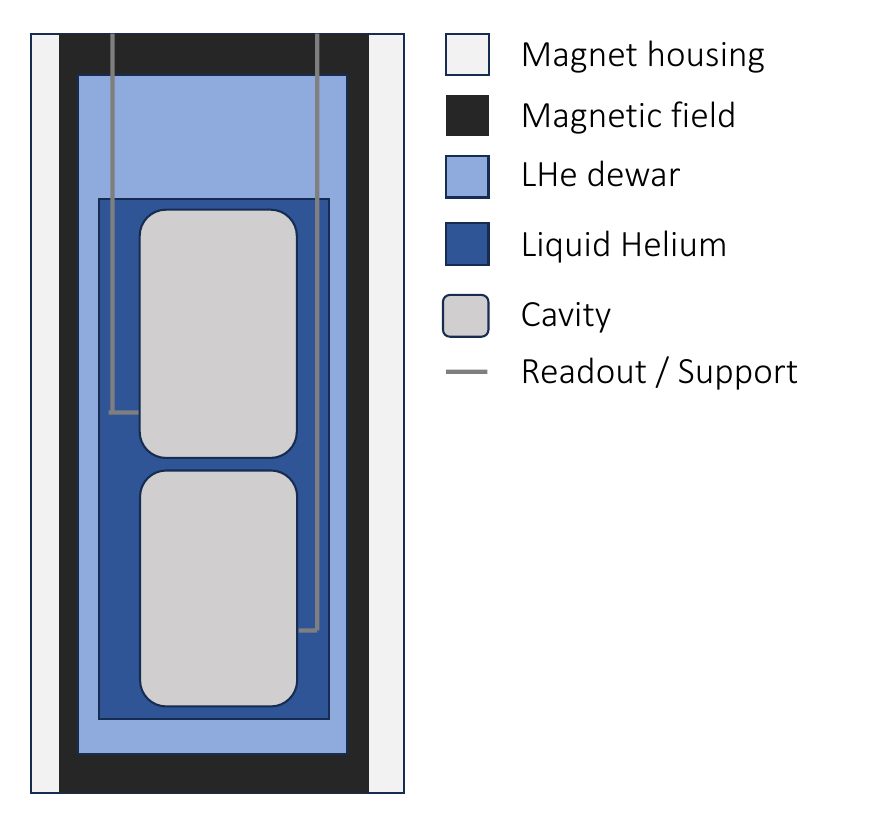}
\caption{Schematic drawing one experimental GravNet-2 setup as single photon experiment.}\label{fig:setup_scheme2}
\end{minipage}
\end{figure*}

\subsection{Possible Extensions}

It should be noted that one could in principle also combine different cavity layouts and magnet systems into GravNet, which are then testing different resonance frequencies at different sensitivities. 
Since such a setup would certainly introduce a certain model dependence in particular when searching for transient signals, its final efficiency still has to be studied in detail. 

\section{GravNet as Photon Counting Experiment (GravNet-2)}
\label{sec:gravNet-2}
\subsection{Setup}

As discussed in Section \ref{sec:SPC}, the shape of the cavity does not increase the likelihood of a conversion, 
but only the active volume of the cavity within the magnetic field is relevant. Given that the cost driving factor is always the magnet system, but not the design of the cavities, we assume the same magnet setup as in GravNet-1 but assume two independent cylindrical cavities with dimensions of $r=4\,cm$ and $h=12\,cm$ instead of three spherical cavities. While the volume increases the sensitivity with $V^{5/3}$, 
one gains significantly more due to the Binomial probabilities, discussed in Section \ref{sec:SPC}.


\subsection{Sensitivities}

Similar to GravNet-1, we assume again N=10 different experimental setups, with a total of $N=20$ operational single and independent cavities, as depicted in Figure \ref{fig:setup_scheme2}.
The cavities operate at a resonance frequency around $5\GHz$ and exhibit a volume of $0.6\,l$ each.
The single RF photon detection efficiency is taken to be $50\%$, a dark count rate of $10\,Hz$ and a time resolution of 0.2 ms are assumed, as discussed in section \ref{sec:concept}, the following sensitivities are expected.

A naive sensitivity estimate yields,
assuming a coincidence time window of 0.2 ms and each setup consisting of 2 independent cavities, a coincidence dark count rate of 1.2 counts per minute.  
Requiring a coincidence of 5 cavities in total a dark count rate of 1 in 190 years is expected. 
The efficiency to detect the coincident production of RF photons in at least 5 out of 20 cavities is calculated using the binomial distribution with $n = 20$ and $p=0.5$ where  $ P(x \ge 5) = 99.4\%$.

The question is however how this translates to a sensitivity on the GW strain $h_0$. 
The photon flux from thermal noise at $0.1$ K and a sensitive bandwidth of 1 kHz is about 10 photons per second at a photon energy of 5 GHz. At 1 MHz sensitive bandwidth would increase the photon flux to 400 Hz. Decreasing the temperature to $0.01$ K would reduces the thermal photon flux by one order of magnitude. 
Hence, we assume for the following calculation a photon flux of 10 Hz from thermal radiation and a negligible contribution to the dark count rate from the detector itself. 
Clearly, to be able to discriminate the thermal noise photons to a signal from a BPH merging event a coincidence measurement is needed, as indicated above.

\begin{figure}[ht]
    \centering
    \includegraphics[width=0.48\textwidth]{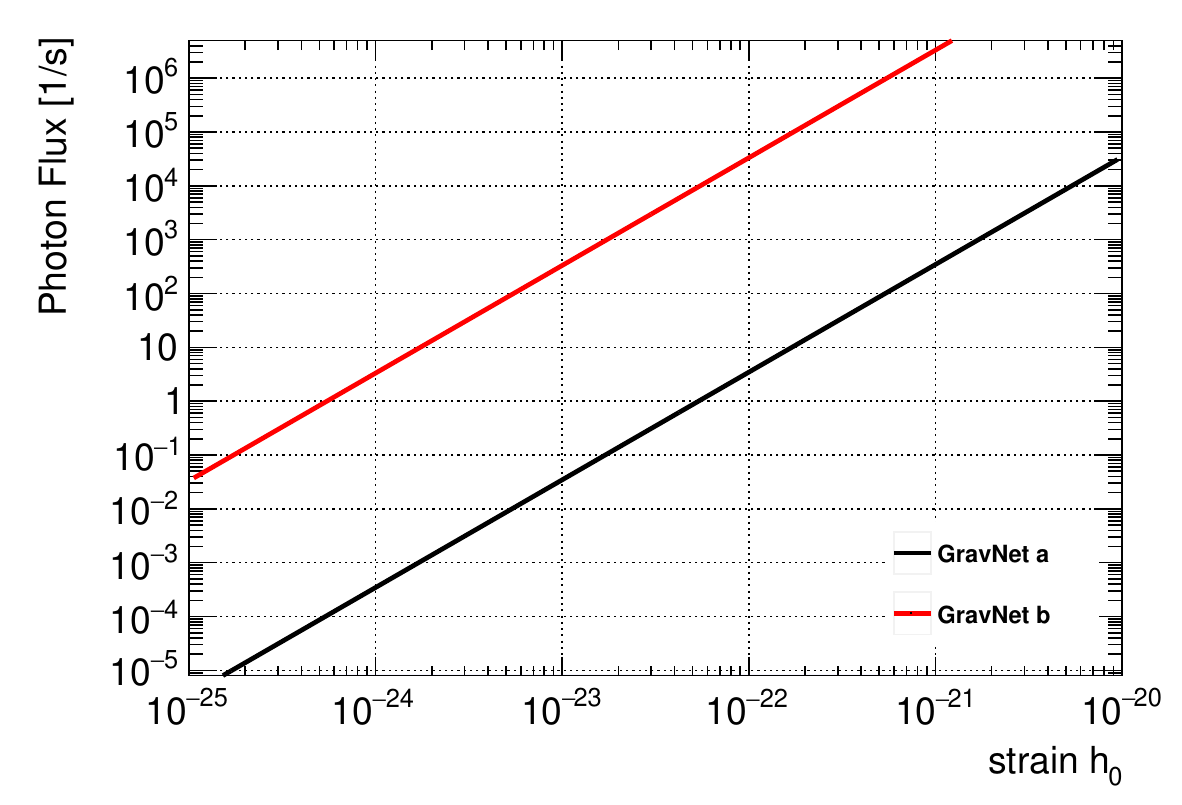}
    \caption{Photon flux in dependence of the GW strain $h_0$ at 5 GHz for GravNet a and b setups detailed in Table \ref{tab:GarvNat2_parameters}.}
    \label{fig:photonFlux}
\end{figure}

The photon flux $\Phi$ generated by a GW can be estimated by dividing the signal power by the photon energy $\Phi = P_{sig} / h\nu$.
Using eq. \ref{eq:signalPower} and assuming $Q_0 = 10^6$ and $\eta = 0.1$ the photon flux generated in one cavity in dependence on the GW strain is shown in Fig. \ref{fig:photonFlux}. Two cavity dimensions are shown: GravNet-a and GravNet-b, whose parameters are summarized in Table \ref{tab:GarvNat2_parameters}.
The smaller cavity (GravNet-a) shows a signal photon flux comparable to the thermal noise of $10\,Hz$ at $h_0 = 1.7\times10^{-21}$ while the larger cavity (GravNet-b) reaches that flux at $h_0 = 1.6\times 10^{-24}$.

\begin{table}[ht]
    \centering
    \begin{tabular}{c|c|c|c}
        Setup &  GravNet-a & GravNet-b \\
        \hline 
        radius   & 40 mm & 40 cm \\
        length  & 12cm & 50 cm \\
        Volume [$m^3$] & $6\times 10^{-4}$ & 0.25 \\
        $Q_0$ &  $10^6$ & $10^5$\\
        $T_{\text{sys}}$ [K]& 0.1 & 0.1 \\
        $B$ [T] & 14 & 9 \\
        \hline
        noise power [W] & $4.4\cdot 10^{-23}\,W$ & $4.4\cdot 10^{-23}\,W$ \\
        $h_0 (P_{\text{sig}} = P_{\text{noise}})$ & $1.6\cdot10^{-22}$ & $3.4\cdot10^{-24}$ & \\
        $\gamma$-flux [1/s] & 10 & 10 & \\
    \end{tabular}
    \caption{Parameters of the experimental setup defining the signal and noise power. The measured values were obtained using the Supax Cu cavity in LHe. The expected values assume a superconducting, spherical cavity with 4 cm radius.}
    \label{tab:GarvNat2_parameters}
\end{table}

\begin{figure*}[ht]
\centering
\begin{minipage}[b]{0.48\textwidth}
\includegraphics[width=0.98\textwidth]{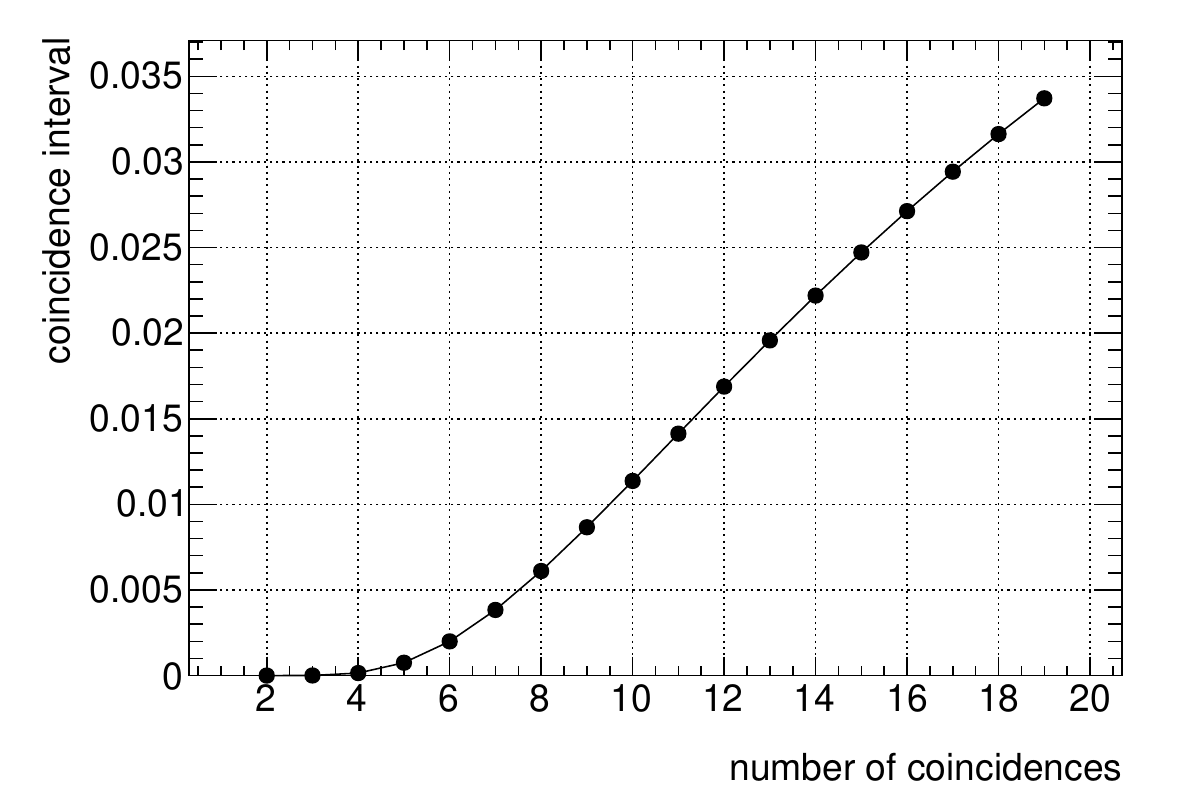}
\caption{Maximum allowed coincidence interval to achieve a background coincidence rate of 1 per year given the number of coincidences and assuming a dark count rate (thermal noise + detector noise) of 10 Hz.}\label{fig:coincidenceInterval}
\end{minipage}
\hspace{0.2cm}
\begin{minipage}[b]{0.48\textwidth}
\includegraphics[width=0.98\textwidth]{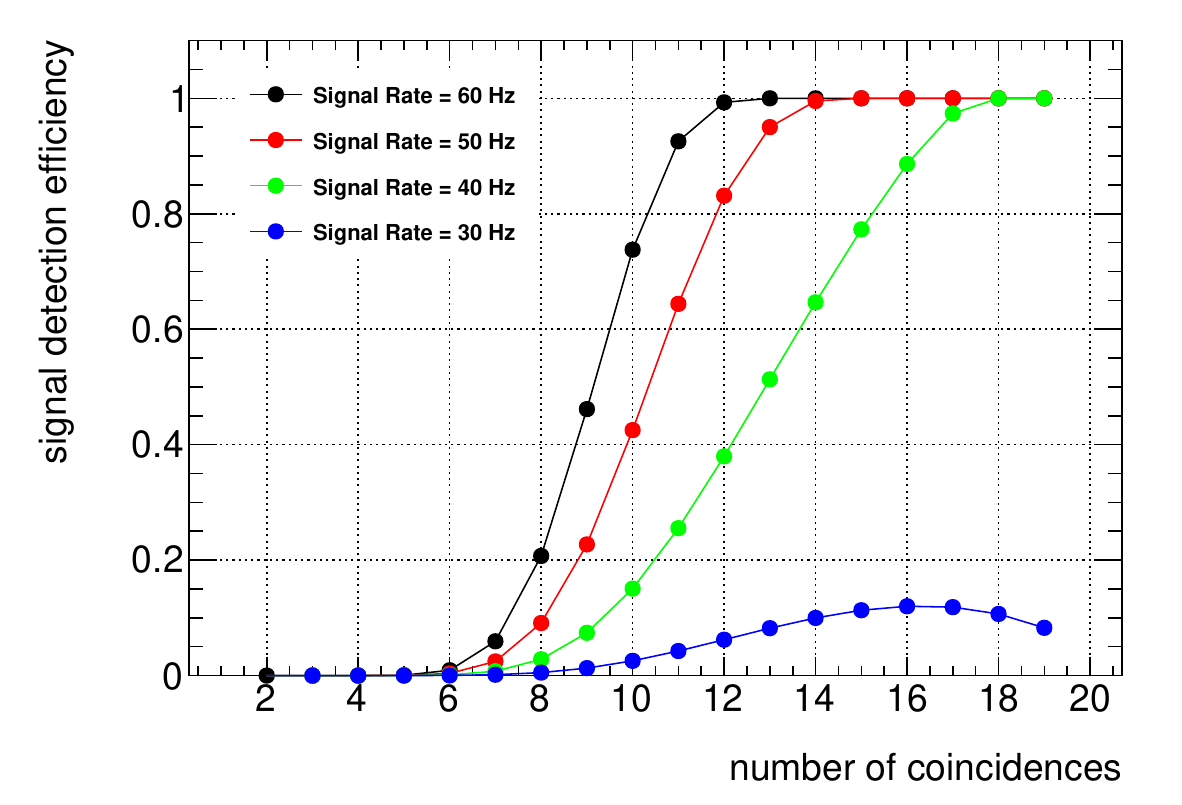}
\caption{Signal efficiency in dependence of the number of coincidences and the coincidence window set according to Figure \ref{fig:coincidenceInterval} for various assumptions on the signal photon rate. A total of 20 independent detectors is assumed.}\label{fig:SignalEfficiency}
\end{minipage}
\end{figure*}

The target rate of accidental coincidences (ac) from the thermal noise are set to one per year. 
This defines the length of the allowed coincidence window $\Delta t$ in dependence on the number of required coincidences $k$ and the background rate $bkg$: 
\[1/\Delta t = (\text{bkg}\cdot \text{secPerYear})^{1/(k-1)} \cdot \text{bkg}\]
This dependence is shown in Figure \ref{fig:coincidenceInterval}. Knowing the needed coincidence interval we can continue and calculate the efficiency to detect one photon from a GW in k detectors within the coincidence window. 
The result is shown for various assumptions on the signal photon flux in Figure \ref{fig:SignalEfficiency}, assuming 20 independent detectors in total. 
A photon flux of $ \Phi = 30\,Hz$ is not reliable detected any more, while for a photon flux of $\Phi = 40\,Hz$ a detection efficiency of 1 is still reached using a coincidence of 18 out of 20 cavities with a coincidence window of $31\,ms$. 
This photon flux corresponds to a GW strain of $h_0 = 3\times10^{-22}$ ($h_0 = 3\times10^{-24}$) for the GravNet-a (GravNet-b) setups, respectively, as is shown in Figure \ref{fig:photonFlux}. 

It is worth stressing the fundamental difference of a counting experiment to the resonance based experiment. 
In the resonant cavity excitation experiment one relies on the signal to be monochromatic w.r.t. to the cavity bandwidth, at least long enough to ring up the cavity so that the maximal RF energy can be read out, which is still below a ms for very high Q cavities. Note that typical integration times for such setups is O(100) seconds, and even in the sensitivity estimate in Section \ref{sec:gravNet-1} 1s integrating time is assumed. 
Then a phase aligned combination of the signals from multiple cavities is required, which may be challenging. 

\begin{figure*}[ht]
\centering
\begin{minipage}[b]{0.68\textwidth}
\centering
\includegraphics[width=0.78\textwidth]{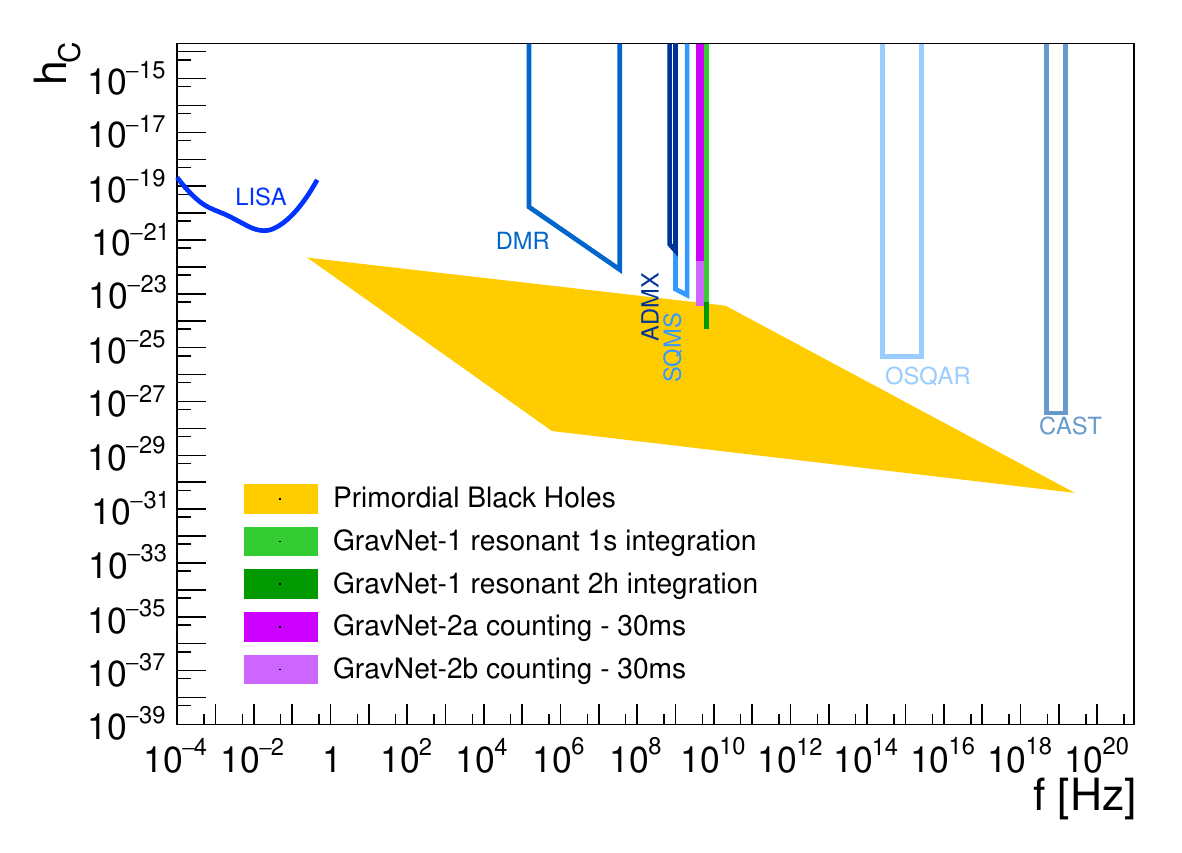}
\caption{Estimated Sensitivity of the GravNet 1 and 2 experiments on the GW strain $h$, assuming 10 setups each. For GravNet-1 two integration times of 1 second and 2 hours are shown. For GravNet-2 sensitivities for two cavity dimensions, a and b as detailed in Table \ref{tab:GarvNat2_parameters}, are shown. The range of PBH merger signals as well as the sensitivity of similar experiments are adapted from \cite{Franciolini:2022htd}. Note that the integration times for the shown sensitivities are much shorter for GravNet than that of typical axion search experiments.}\label{fig:results}
\end{minipage}
\end{figure*}

Contrary, in the counting experiment we rely on simultaneous measurements of single photons within a time interval of $ms$ making the experiment sensitive to very short, transient signals. The cavity signals don't need to be phase aligned and even measurements from setups with variations in frequencies can easily be combined.
Using order of 20 individual detectors the experiment becomes essentially background free given the parameters of the detectors as stated above, 
and can be sensitive to strains down to $10^{-22}$ to $10^{-24}$, depending on the size of the conversion volume. Those results are summarized in Figure \ref{fig:results} in comparison to other existing experiments. 

\section{Conclusion}

In this paper we propose to setup a world-wide network of cavity based experiments to search for high frequency gravitational waves (GravNet) with frequencies in the 5 GHz regime. 
Assuming ten participating laboratories, sensitivities on the gravitational wave strain of $h_0 < 10^{-24}$ for monochromatic sources and long integration times and $h_0 < 10^{-22}$ for $~10\,ms$ transient signals can be reached, depending on the chosen experimental setup. 
Using larger magnet setups with lower field would even extend the sensitive to transient signals to $h_0 < 10^{-24}$. This would allow for a first search of several exotic sources of high frequency gravitational waves as well as for primordial black hole models with masses larger than $3\times 10^{-7}\,M_\odot$.  
In particular it should be noted, that the experimental concept is based on commercial magnet systems, which are rather cost efficient compared to specialized dedicated high field magnets. Moreover, a  worldwide collaboration such as GravNet would share costs automatically with local lab-based experiments and boost the R\&D effort within many groups, since new developments at one site can be easily integrated in all other experimental locations.

This paper should be seen as a starting point for discussions towards a collaborative world-wide effort in the context of gravitational wave physics beyond the regime of classic astrophysical objects.


\section*{Acknowledgement}
We thank Pedro Schwaller, Sebastian Schenk and Tim Schneemann for the helpful discussions and comments during the preparation of this paper. This work would have not been possible without the continuous support from the PRISMA+ Cluster of Excellence at the University of Mainz.  
\bibliographystyle{apsrev4-1} 
\bibliography{./GravNet}
	
\end{document}